\documentclass[11pt]{article}
\usepackage{amssymb}
\usepackage{epsfig}
\begin{document}
\title{{\bf{\Large Connection between response parameter and anomaly coefficient in two dimensional
 anomalous fluid}}}
\author{ 
 {\bf {\normalsize Bibhas Ranjan Majhi}$
$\thanks{E-mail: bibhas.majhi@mail.huji.ac.il}}\\ 
{\normalsize Racah Institute of Physics, Hebrew University of Jerusalem,}
\\{\normalsize Givat Ram, Jerusalem 91904, Israel}
\\[0.3cm]
}
\maketitle

\begin{abstract}
In ($1+1$) dimensional hydrodynamics in presence of the gravitational anomalies, the constitutive relations for the stress tensor contain the response
 parameters $\bar{C}_1$, $\bar{C}_2$ and the gravitation anomaly coefficients $c_g$, $c_w$. Here it is shown that
they are related by the two relations $\bar{C}_1 = 4\pi^2 c_w$ and $\bar{C}_2 = 8\pi^2 c_g$. This agrees with the earlier findings.
I argue that the {\it Israel-Hartle-Hawking vacuum} 
is the natural boundary condition which leads to such relation. Finally, the possible physical 
implications are discussed.
\end{abstract}

\section{Introduction}
The relevant equation of motion for the hydrodynamics are the certain conservation equations which
are supplemented by the constitutive relations -- the energy-momentum tensor and the current expressed 
in terms of the fluid variables \cite{Landau}.
The role of anomalies in fluid dynamics has been a central attention in recent years 
\cite{Dubovsky:2011sk} -- \cite{Banerjee:2013fqa}. The central purpose, so far, is to find the 
constitutive relations by solving the anomaly expressions. It turns out that 
the these relations can be expressed in terms of fluid variables, like temperature, chemical 
potential etc. under the static background. This is possible because the fluid variables at an arbitrary
 spacetime point are related to their equilibrium values through the background metric coefficients
 which is known as the Tolman relation \cite{Tolman}.

   The constitutive relations for the stress tensor in presence of the gravitational 
anomalies usually contain certain factors, like response parameters and the coefficients 
of the anomaly expressions (For details, see \cite{Valle:2012em} -- \cite{Banerjee:2013fqa}). In this 
paper, I will show that there exists two simple 
relations between them. It agrees with the earlier findings \cite{Jensen:2012kj}. Contrary to the existing work \cite{Jensen:2012kj}, the present 
analysis will be simple and will provide more physical insight in this paradigm. It will be 
shown that the imposition of the {\it Israel-Hartle-Hawking vacuum} condition on the two dimensional 
anomalous stress tensor yields the required result. 

Here the detail derivation of anomalous 
constitutive relation will not be given, because it has been already done earlier 
\cite{Banerjee:2013qha,Banerjee:2013fqa}. To achieve the goal, I will start from the anomalous 
constitutive relations which are derived in \cite{Banerjee:2013fqa} and then proceed further. 
 At the end, the physical implications will be discussed. 

The organization of the paper is as follows. In section \ref{summary}, the form of the anomalous 
constitutive relations for the energy-momentum tensor will be given. Then I shall find the relations between 
the response parameters and the anomaly coefficients in section \ref{relationpc}. Final section 
will be devoted for the conclusions and physical interpretations.

\section{\label{summary}Two dimensional anomalous fluid: a brief summary}
   In this section, the explicit form of the constitutive relations for the energy-momentum tensor 
in presence of the gravitational anomalies in ($1+1$) dimensions will be given. The background is chosen 
to be static. Here, I will not give the detail derivation of these relations since it can be followed from
\cite{Banerjee:2013fqa}. 

 It is well known that the anomalies are the intrinsic properties of field theory. In ($1+1$)  
the anomaly expressions are given by \cite{Jensen:2012kj} (See also \cite{Bardeen:1984pm} -- \cite{Banerjee:2008wq}): 
\begin{equation}
\nabla_b T^{ab} =  c_g\bar{\epsilon}^{ab}\nabla_b R; \,\,\,\ T^{a(g)}_{a} = c_w{R}~,
\label{diffchiral}
\end{equation}
where $c_g$ and $c_w$ are two normalization constants and $R$ is the two dimensional Ricci scalar. 
Following the steps employed in \cite{Banerjee:2013qha,Banerjee:2013fqa}, it can be shown that in the comoving frame under the static background: 
\begin{equation}
ds^2 = - e^{2\sigma(r)}dt^2 + g_{11} dr^2~,
\label{effectivemetric}
\end{equation}
the components of stress tensor, obtained from the solutions of the anomaly 
expressions Eq. (\ref{diffchiral}), can be casted in the constitutive relation for stress tensor of a fluid:
\begin{eqnarray}
\label{chi2}
T_{ab} &=& \left[2c_w\left(u^c \nabla^d - u^d\nabla^c\right)\nabla_c u_d + 2\bar{C}_1T^2 \right]{u}_a{u}_b
\nonumber
\\
&-& \left[2c_g\left(u^c \nabla^d - u^d\nabla^c\right)\nabla_c u_d + \bar{C}_2T^2 \right]({u}_a\tilde{u}_b+\tilde{u}_au_b)
\nonumber
\\
&+& \left[\bar{C}_1T^2-c_w\left(u^c \nabla^d \nabla_d u_c \right)\right]g_{ab}~.
\end{eqnarray}
Here $u^a$ is the fluid velocity,  $\tilde{u}_a = \bar{\epsilon}_{ab}u^b$ and $T$ is related to the equilibrium temperature $T_0$ by the relation $T = e^{-\sigma}T_0$. 
$u^a$ is chosen such that in the comoving frame it satisfies the timelike 
condition $g_{ab}u^au^b = -1$. For a detail analysis and the meaning of the symbols, see \cite{Banerjee:2013fqa}. 
The similar relation was also obtained by derivative expansion approach in \cite{Jensen:2012kj}. The 
response parameters $\bar{C}_1$ and $\bar{C}_2$ are related to the equilibrium temperature $T_0$ of the fluid as,
\begin{equation}
\bar{C}_1=(C_{uu}+C_{vv}){T_0}^{-2}; \,\,\,\ \bar{C}_2=(C_{uu}-C_{vv}){T_0}^{-2}
\label{cbar}
\end{equation}
where $C_{uu}$ and $C_{vv}$ are the integration constants appearing in the solution 
of Eq. (\ref{diffchiral}).

  Now it must be noted that in the above constitutive relation Eq. (\ref{chi2}) 
there exists four parameters: two response parameters $\bar{C}_1$ and $\bar{C}_2$ and two anomaly coefficients $c_g$ and $c_w$. In the next 
section I will provide a relation between them.

\section{\label{relationpc}Relation between response parameters and anomaly coefficients}
  In the derivative expansion approach \cite{Jensen:2012kj}, the authors 
of the paper showed that the response parameters $\bar{C}_1$ and $\bar{C}_2$ are proportional to the normalization 
constants $c_w$ and $c_g$ appearing in the anomaly expressions, respectively. The exact connection is given by,
\begin{equation}
\bar{C}_1 = 4\pi^2 c_w~; \,\,\,\ \bar{C}_2 = 8\pi^2 c_g~.
\label{relation}
\end{equation} 
(See, Eq. (4.15a) and Eq. (4.15b) of \cite{Jensen:2012kj} with the identifications $\bar{C}_1 = p_0$ and $\bar{C}_2 = -\tilde{c}_{2d}$). 
The evidence of this statement can be followed from  earlier works in various 
subfield of physics \cite{Volovik:1999wx} -- \cite{Kraus:2005zm}. The derivation in \cite{Jensen:2012kj} 
 was done by demanding that the thermal energy-momentum 
tensor on the cone must vanishes in the Euclidean vacuum. The procedure is very much technically
involved. 
Here, in this paper, a simple realization of this relation will be 
presented. I shall show that one can understand it by just fixing the integration constants $C_{uu}$ and 
$C_{vv}$ upon imposing a relevant boundary condition.

  To proceed towards the main goal, let us first express the components of $T_{ab}$ 
in terms of the metric coefficients under the background (\ref{effectivemetric}) in null 
coordinates ($u,v$). From Eq. (\ref{chi2}), these are given by,
\begin{eqnarray}
&&T_{uu} = \frac{2c_g+c_w}{4}\frac{e^{2\sigma}}{g_{11}^2} (2\sigma''g_{11} - \sigma'g'_{11}) + C_{uu}~; 
\label{uu}
\\
&&T_{vv} = -\frac{2c_g-c_w}{4}\frac{e^{2\sigma}}{g_{11}^2} (2\sigma''g_{11} - \sigma'g'_{11}) + C_{vv}~;
\label{vv}
\\ 
&&T_{uv} =-\frac{c_w}{4} e^{2\sigma}R~.
\label{uv}
\end{eqnarray}
Here the prime denotes the derivative with respect to ``$r$'' coordinate.
Now to determine the integration constants $C_{uu}$ and $C_{vv}$ we need to impose a suitable boundary condition. In 
literature there exists three types of vacua, corresponding to the different boundary conditions \cite{Balbinot:1999vg}.
 Among them, here we shall show that {\it Israel-Hartle-Hawking vacuum} condition is the relevant one to achieve the 
required relation.
 This vacuum is defined by the fact that the stress tensor in Kruskal coordinates
 corresponding to the both outgoing and ingoing modes must be regular 
 near the horizon. 
Hence $T_{uu}\rightarrow 0$ and $T_{vv}\rightarrow 0$ near the horizon. To impose 
these conditions, let us consider that the metric (\ref{effectivemetric}) is the solution of the 
Einstein equation and since the metric is static, the Killing horizon and the event horizon must 
coincide \cite{Carter}. Therefore, in the present case, 
$e^{2\sigma}|_{r_0} =0 = 1/g_{11}|_{r_0}$ where $r_0$ is the position of the horizon. Now denoting 
$e^{2\sigma(r)}\equiv f(r)$ and $1/g_{11}(r)\equiv g(r)$, we expand them the near the horizon as
\begin{eqnarray}
f(r) = f'(r_0)(r-r_0) + \dots ; \,\,\,\,\ g(r) = g'(r_0)(r-r_0)+\dots
\label{nearhorioncoff}
\end{eqnarray}
Substituting these in the $uu$ component of $T_{ab}$ (see Eq. (\ref{uu})) and then taking
 the limit $r\rightarrow r_0$, we obtain 
expression as
\begin{equation}
T_{uu} = -\frac{2c_g+c_w}{4}\frac{f'(r_0)g'(r_0)}{2}+C_{uu}~.
\label{nearuu} 
\end{equation}
Now since near horizon this must vanishes, we get 
\begin{equation}
C_{uu} = \frac{(2c_g+c_w)\kappa^2}{2} = 2\pi^2(2c_g+c_w) T_0^2~,
\end{equation}
 where the surface gravity $\kappa = \sqrt{f'(r_0)g'(r_0)}/2$ and the equilibrium temperature $T_0 = \kappa/2\pi$. 
Similarly, imposition of the same boundary condition on Eq. (\ref{vv}) leads to
\begin{equation}
C_{vv} =  -2\pi^2(2c_g-c_w) T_0^2
\end{equation}
Substitution of these in Eq. (\ref{cbar}) lead to both the two required relations in Eq. (\ref{relation}).

   Before I conclude, let me make some comments on the relations of Eq. (\ref{relation}) and the way these have been derived here. 
In the context of ($1+1$) dimensional conformal field theory (CFT) and Cardy formula, there exits similar results like Eq. (\ref{relation}). 
It is well known that the Cardy formula relates the pressure of the CFT with the left and right handed central charges 
by the relation $P = 2\pi T^2 (c_L+c_R)/24 = 4\pi^2 c_w T^2$ where $c_w$ is the same trace anomaly coefficient as given in Eq. (\ref{diffchiral})
 (See \cite{Bloete:1986qm}, for details). Furthermore, the arguments employed here to derive the relations are similar to those used for deriving the Cardy formula. 
Here we imposed the Israel-Hartle-Hawking vacuum condition which for Killing horizons implies that the stress-tensor must be regular on 
the corresponding Euclidean cigar. Moreover, it is possible to derive the relations by using the similar formalism as that of the derivation of 
the Cardy formula. This has been shown in \cite{Jensen:2012kj}. On the other hand if one calculates the $T^r_r$ component, for the present case, which is the pressure, 
then it turns out to be $T^r_r=4\pi^2c_w T^2+$ (higher derivative terms). 
This is exactly the expression, mentioned above, obtained by Cardy formula. In addition, it can be shown that $T^r_t = 8\pi^2 T^2 c_g$ and $-T^t_t = 4\pi^2 c_w T^2$ upto some higher derivative terms. All these suggest that the method, employed here, is similar to that for deriving the Cardy formula.

\section{Conclusions}
  In this paper, a simple derivation of the relations in (\ref{relation}) has been given in the context of 
two dimensional anomalous hydrodynamics. I showed that
 these are coming from the fixing of the integration constants, appearing in the solutions of the 
gravitational anomaly equations, by imposing the {\it Israel-Hartle-Hawking vacuum} condition. Such an analysis, 
so far I know, does not exist in literature. 
It clearly revels the importance of the vacuum condition in the anomalous hydrodynamics. 
I believe that the present analysis might shed some light towards this paradigm.

   The physical significance is as follows. It is well known that in the thermodynamics of 
gravity, the main macroscopic 
entities are temperature, entropy, free energy, etc. which are all observer dependent quantities. 
For instance, a freely falling observer through the black hole horizon does not associate these on the 
horizon while the thermodynamical parameters are well defined with respect to the outside static observer.
 Therefore, one can argue that the degrees of freedom responsible for them are not absolute 
\cite{Majhi:2012tf} -- \cite{Majhi:2013jpk}. Similarly, 
as we have seen in the present analysis, the relevant observer is one which corresponds to the Israel-Hartle-Hawking vacuum. 
Hence {\it the underlying microscopic theory for the anomalous fluid dynamics may have observer dependent 
notion}. Finally, the present analysis is general enough to include higher dimensional
 theories. Furthermore, to understand more about the significance of the vacuum it would be interesting to 
study the gravitation anomalies in higher dimensions. This I leave for the future.
    
\vskip 5mm
\noindent
{\bf{Acknowledgements}}\\
\noindent
I thank Rabin Banerjee and S. Jain for several useful discussions. I also like to thank the referee for various important comments. 
The research of the author is supported by a Lady Davis Fellowship at Hebrew University, 
by the I-CORE Program of the Planning and Budgeting Committee and the Israel Science Foundation 
(grant No. 1937/12), as well as by the Israel Science Foundation personal grant No. 24/12.


\begin{thebibliography}{99}

\bibitem{Landau}
  L.~D.~Landau, E.~M.~Lifshitz,
  ``Fluid Mechanics (Course of Theoretical Physics - volume 6),''
  Elsevier, UK.

  
\bibitem{Dubovsky:2011sk} 
  S.~Dubovsky, L.~Hui, A.~Nicolis,
  ``Effective field theory for hydrodynamics: Wess-Zumino term and anomalies in two spacetime dimensions,''
  arXiv:1107.0732 [hep-th].
  
\bibitem{Valle:2012em} 
  M.~Valle,
  ``Hydrodynamics in 1+1 dimensions with gravitational anomalies,''
  JHEP {\bf 1208}, 113 (2012)
  [arXiv:1206.1538 [hep-th]].
   
\bibitem{Jensen:2012kj} 
  K.~Jensen, R.~Loganayagam, A.~Yarom,
  ``Thermodynamics, gravitational anomalies and cones,''
  JHEP {\bf 1302}, 088 (2013)
  [arXiv:1207.5824 [hep-th]].
  
  \bibitem{Jains:2013}
  S.~Jain, T.~Sharma,
 	 ``Anomalous charged fluids in 1+1d from equilibrium partition function,''  
 JHEP {\bf 1301}, 039 (2013)
  [arXiv:1203:5308 [hep-th]].
 
 \bibitem{Banerjee:2013qha} 
  R.~Banerjee,
  ``Exact results in two dimensional chiral hydrodynamics with diffeomorphism and conformal anomalies,''
  arXiv:1303.5593 [gr-qc].

 
\bibitem{Banerjee:2013fqa} 
  R.~Banerjee, P.~Chakraborty, S.~Dey, B.~R.~Majhi and A.~K.~Mitra,
  ``Two dimensional hydrodynamics with gauge and gravitational anomalies,''
  arXiv:1307.1313 [hep-th].

\bibitem{Tolman}
 R.~C.~Tolman, 
 “Relativity, Thermodynamics and Cosmology”, New York: Dover Publication,
 (1987), p.318.

 
\bibitem{Bardeen:1984pm} 
  W.~A.~Bardeen, B.~Zumino,
  ``Consistent and Covariant Anomalies in Gauge and Gravitational Theories,''
  Nucl.\ Phys.\ B {\bf 244}, 421 (1984).
   
 
  \bibitem{Gaume-Gins:1985} 
     L.~A.~Gaume, P.~Ginsparg
     ``The structure of gauge and gravitational anomalies,''
  Annals\ of\ Phys. {\bf 161}, 423 (1985)
 
  
  \bibitem{Gaume-witten:1984} 
  L.~A.~Gaume, E.~Witten,
  ``Gravitational anomalies''
  Nucl.\ Phys.\ B {\bf 234}, 269 (1984).

\bibitem{Banerjee:2008wq} 
  R.~Banerjee and S.~Kulkarni,
  ``Hawking Radiation, Covariant Boundary Conditions and Vacuum States,''
  Phys.\ Rev.\ D {\bf 79}, 084035 (2009)
  [arXiv:0810.5683 [hep-th]].\\
  R.~Banerjee and B.~R.~Majhi,
  ``Connecting anomaly and tunneling methods for Hawking effect through chirality,''
  Phys.\ Rev.\ D {\bf 79}, 064024 (2009)
  [arXiv:0812.0497 [hep-th]].


\bibitem{Volovik:1999wx} 
  G.~E.~Volovik and A.~Vilenkin,
  ``Macroscopic parity violating effects and He-3-A,''
  Phys.\ Rev.\ D {\bf 62}, 025014 (2000)
  [hep-ph/9905460].

\bibitem{Cappelli}
  A.~ Cappelli, M.~ Huerta, G.~ R.~ Zemba,
  ``Thermal Transport in Chiral Conformal Theories and Hierarchical Quantum Hall States,''
  Nucl.\ Phys.\ B {\bf 636}, 568 (2002)
  [cond-mat/0111437].

\bibitem{Stone}
  M.~ Stone,
  ``Gravitational Anomalies and Thermal Hall effect in Topological Insulators,''
  Phys.\ Rev. \ B {\bf 85}, 184503 (2012)
  [arXiv:1201.4095]

\bibitem{Landsteiner:2011cp} 
  K.~Landsteiner, E.~Megias and F.~Pena-Benitez,
  ``Gravitational Anomaly and Transport,''
  Phys.\ Rev.\ Lett.\  {\bf 107}, 021601 (2011)
  [arXiv:1103.5006 [hep-ph]].

\bibitem{Kraus:2005zm} 
  P.~Kraus and F.~Larsen,
  ``Holographic gravitational anomalies,''
  JHEP {\bf 0601}, 022 (2006)
  [hep-th/0508218].

\bibitem{Balbinot:1999vg} 
  R.~Balbinot, A.~Fabbri and I.~L.~Shapiro,
  ``Vacuum polarization in Schwarzschild space-time by anomaly induced effective actions,''
  Nucl.\ Phys.\ B {\bf 559}, 301 (1999)
  [hep-th/9904162].

\bibitem{Carter}
  B.~ Carter, 
  “Black Hole Equilibrium States” in Black Holes, ed. by
C. DeWitt and B.S. DeWitt, 57-214, Gordon and Breach (New York,
1973).

\bibitem{Bloete:1986qm} 
  H.~W.~J.~Bloete, J.~L.~Cardy and M.~P.~Nightingale,
  ``Conformal Invariance, the Central Charge, and Universal Finite Size Amplitudes at Criticality,''
  Phys.\ Rev.\ Lett.\  {\bf 56}, 742 (1986).\\
  I.~Affleck,
  ``Universal Term in the Free Energy at a Critical Point and the Conformal Anomaly,''
  Phys.\ Rev.\ Lett.\  {\bf 56}, 746 (1986).

\bibitem{Majhi:2012tf} 
  B.~R.~Majhi, T.~Padmanabhan,
  ``Noether current from the surface term of gravitational action, Virasoro algebra and horizon entropy,''
  Phys.\ Rev.\ D {\bf 86}, 101501 (2012)
  [arXiv:1204.1422 [gr-qc]].
  

\bibitem{Majhi:2012nq} 
  B.~R.~Majhi,
  ``Noether current of the surface term of Einstein-Hilbert action, Virasoro algebra and entropy,''
  Adv.\ High Energy Phys.\  {\bf 2013}, 386342 (2013)
  [arXiv:1210.6736 [gr-qc]].


\bibitem{Majhi:2013jpk} 
  B.~R.~Majhi and T.~Padmanabhan,
  ``Thermality and Heat Content of horizons from infinitesimal coordinate transformations,''
  Eur.\  Phys.\  J.\  C {\bf 73}, 2651 (2013)
  [arXiv:1302.1206 [gr-qc]].


\end{thebibliography}
\end{document}